\documentclass[conference]{IEEEtran}
\IEEEoverridecommandlockouts
\usepackage{amsmath,amssymb,amsfonts}
\usepackage{algorithm}
\usepackage{algpseudocode}
\usepackage{array}
\usepackage{textcomp}
\usepackage{stfloats}
\usepackage{url}
\usepackage{color}
\usepackage{verbatim}
\usepackage{graphicx}
\usepackage{tabularx}
\usepackage{multirow}
\usepackage{array}
\usepackage{booktabs}
\usepackage{graphicx}
\usepackage{parskip}

\usepackage[numbers]{natbib}
\usepackage{bm}
\usepackage{mathtools}

\usepackage{subfigure}
\usepackage{mathrsfs}

\usepackage{bibspacing}
\graphicspath{ {./fig/} }
\usepackage{enumitem}
\def\BibTeX{{\rm B\kern-.05em{\sc i\kern-.025em b}\kern-.08em
    T\kern-.1667em\lower.7ex\hbox{E}\kern-.125emX}}
\begin{document}

\title{Spatiotemporal Trust Evaluation for Collaborator Selection via Customized GNN-Mamba} 

\author{\IEEEauthorblockN{Botao Zhu and Xianbin Wang}
\IEEEauthorblockA{Dept. of Electrical and Computer Engineering, Western University,
London, Ontario N6A 3K7 CANADA \\}}




\maketitle

\begin{abstract}
  The successful completion of collaborative tasks relies on the effective selection of trustworthy collaborators. To accurately evaluate the trustworthiness of potential collaborators, it is necessary to combine insights from their past collaborations with assessments of their resource capabilities under specific task contexts. However, the coexistence of diverse trust perspectives, along with complex spatiotemporal dependencies among devices, makes accurate trust evaluation particularly challenging. To address these challenges, we propose a customized Graph Neural Network (GNN)-Mamba (GM) model for trust evaluation and collaborator selection. In this model, the GNN model performs spatial trust fusion by leveraging inter-device spatial dependencies extracted from historical collaborations, while the Mamba-based temporal model captures both short-term fluctuations and long-term evolution of device trust. In addition, task-specific resource trust is incorporated to reflect the practical capabilities of devices under varying task conditions. Experimental results demonstrate that the proposed GM model outperforms baseline approaches in terms of the accuracy and stability of trust evaluation.
\end{abstract}


\section{Introduction}
Modern applications and interconnected systems often involve computation-heavy tasks that exceed the capabilities of individual devices, which are constrained by limited processing power and energy. Distributed resource scheduling provides a potential solution by enabling offloading of tasks to networked devices with higher computational capacity. For example, in UAV swarms, drones can offload real-time image processing or navigation computations to a leader drone or edge server to conserve onboard resources~\cite{9892691}, whereas in healthcare IoT networks, wearable devices often offload patient data analysis to local hospital servers to maintain real-time monitoring~\cite{10288131}. These examples underscore the importance of choosing reliable collaborators to guarantee task completion.

Trust has emerged as a critical indicator of device reliability, reflecting confidence in a collaborator’s ability to perform specific tasks~\cite{11096939}. Various methods have been proposed recently to evaluate the trustworthiness of collaborators. Some studies assess trust based on historical collaboration performance--utilizing metrics such as packet loss rate, response time, and task success rate to determine past reliability of devices. Others evaluate trust from the perspective of resource availability, employing indicators such as energy level, bandwidth, or CPU utilization to characterize a device’s trustworthiness~\cite{Najib2024QSTrust}. However, these approaches face notable limitations. First, they rely on generic resource indicators to evaluate device resource trust, overlooking the fact that identical resources may exhibit different levels of trustworthiness across distinct task contexts. Second, they fail to capture the spatiotemporal trust dependencies embedded in historical collaborations. Devices involved in the same collaboration have spatial trust correlations, whereas a device’s trustworthiness evolves over time as it engages in collaborations with different entities. Therefore, accurate evaluation of device trustworthiness requires a customized trust evaluation model. Such a model should not only assess historical trust by capturing the spatiotemporal dependencies among devices from past collaborations, but also enable task-specific evaluation of device resources.

However, capturing trust dependencies across both temporal and spatial dimensions remains challenging. In the spatial dimension, due to the complexity of inter-device collaborations, a device's trustworthiness involves not only evaluations from its direct collaborators but also recommendations from indirect neighbours. Traditional approaches, such as simple weighting and rule-based assessments, are often inadequate for modeling such complex spatial relationships. Recently, Graph Neural Networks (GNNs) have been applied to model intricate spatial interactions in social IoT networks~\cite{9488814}. By representing participants as nodes and their interactions as edges, GNNs effectively encode spatial correlations among nodes. Through the iterative message-passing process of GNNs, each node aggregates information from both direct and multi-hop neighbours. Due to their ability to capture complex spatial relationships, GNNs are suitable for modeling spatial trust dependencies among devices in this study.

In the temporal dimension, a device's trustworthiness is influenced not only by recent collaborations but also by long-term collaborations, necessitating the integration of both short-term and long-term collaborations. Existing temporal fusion methods, such as Recurrent Neural Networks (RNNs), are largely limited to short-term dependencies, making them unsuitable for capturing long-range temporal dynamics~\cite{Alghofaili2022TrustManagement}. Conversely, Transformer-based models offer powerful long-range dependency modeling, but their self-attention mechanism incurs quadratic computational complexity. Mamba is a deep neural sequence model introduced in 2023 that builds on state‑space modeling ideas to efficiently handle very long input sequences~\cite{gu2023mamba}. Unlike classical structured state space models with fixed dynamics, Mamba introduces input‑dependent parameters that allow the model to adjust how much past information is propagated or suppressed at each step, effectively behaving as a selective state space model (SSM). By making key transition and readout parameters functions of the current input, Mamba gains the ability to focus processing on the most contextually relevant parts of a sequence, similar in spirit to attention mechanisms but without the associated quadratic complexity in sequence length. Therefore, it serves as a promising tool for modeling trust dependencies along the temporal dimension in this study.


Leveraging the significant advantages of GNNs and Mamba, we propose a novel spatiotemporal trust evaluation model, GNN-Mamba (GM), for collaborator selection. Historical collaboration records are first converted into a series of graph snapshots, capturing relationships among devices at different time intervals. In the spatial domain, a GNN module aggregates trust information from each device’s direct neighbors as well as multi-hop connections, effectively modeling complex inter-device dependencies. For the temporal dimension, we employ a Mamba-based mechanism to integrate trust information across the sequence of historical snapshots. 
Finally, the framework incorporates task-specific resource evaluation to measure each candidate device’s suitability for the given task. The proposed GM model fully leverages spatiotemporal information and task-specific features, enabling precise and context-aware assessment of collaborator trustworthiness. The main contributions of this paper are summarized as follows.
\begin{itemize}[leftmargin=*]
    \item We propose a customized spatiotemporal trust model that leverages the spatiotemporal dependencies among devices reflected in historical collaboration data and the task-specific characteristics of device resources, enabling accurate and context-aware evaluation of device trustworthiness for reliable collaborator selection. 

    \item  We develop a GNN-aided spatial trust fusion mechanism that effectively encodes and captures the complex, high-order spatial dependencies embedded in devices’ historical collaborations.

    \item We innovatively leverage the Mamba architecture to design a temporal trust fusion mechanism to accurately capture the dynamic evolution of device trust.
\end{itemize}

\section{System Model and Problem Description}
We consider a collaborative system comprising a set of devices $\bm{A} = \{a_1,\dots,a_I\}$.
Each device can act either as a task owner, generating tasks, or as a collaborator, executing tasks initiated by others. A monitoring device is deployed within the system to monitor collaborative interactions and collect performance metrics from collaborators. The collected collaboration records over a long time period $\mathbb{T}$ are represented by a dataset $\bm{F}$. Each record $f_{(a_i,a_j)} \in \bm{F}$ represents a collaboration event in which device $a_j$ assists device $a_i$ in completing a task. It includes performance metrics of device $a_j$, such as packet loss rate, task completion outcomes, and other relevant metrics. We assume that device $a_i$ generates a task $\beta = (\beta^{\text{size}}, \beta^{\text{des}}, \beta^{\text{time}})$, where $\beta^{\text{size}}$ denotes the task size, $\beta^{\text{des}}$ represents the processing density (cycles/bit), and $\beta^{\text{time}}$ indicates the maximum tolerable completion time. Due to limited computational resources, device $a_i$ offloads the task $\beta$ to a reliable collaborator for execution. Accurately evaluating the trustworthiness of potential collaborators, considering both their historical collaboration performance and available resources, is therefore essential. Formally, the trustworthiness of device $a_j$ as assessed by device $a_i$ is defined as:
\vspace{-0.05 in}
\begin{align}
    T_{(a_i,a_j)}  =  T^{\text{his}}_{(a_i,a_j)}(\bm{F})T^{\text{res}}_{(a_i,a_j)}(\beta),
\end{align}
where $T^{\text{his}}_{(a_i,a_j)} \in [0,1]$ quantifies the historical trustworthiness of device $a_j$, and $T^{\text{res}}_{(a_i,a_j)} \in [0,1]$ measures the extent to which device $a_j$’s available resources meet the requirements of the task $\beta$.
The resources of a device comprise both communication and computation capabilities, including CPU capacity, storage space, collaboration willingness, available time, and so on, which collectively determine its ability to participate in task execution. After evaluating all potential collaborators, device $a_i$ selects the one with the highest trust value as its final collaborator for task execution.

To achieve accurate trust evaluation, a customized assessment approach is required, leveraging the characteristics of both historical collaboration data and device resources. Historical collaboration data inherently exhibit strong spatiotemporal properties. Temporally, a device’s trust can vary across different time slots, performing reliably during certain periods while exhibiting instability at others. Spatially, trust can only be evaluated among devices that have previously collaborated.

\section{Preliminary} 

In this section, we introduce the fundamental principles of GNNs and Mamba.
\subsection{GNNs} 
In GNNs, the representation of each node is determined not only by its own features but also by the features of its neighbouring nodes. GNNs learn node embeddings through a message propagation and aggregation mechanism, allowing each node to capture relational dependencies within the graph structure. At the $l$-th layer, each neighbour of node $a$ propagates its message to node $a$: 
\begin{align}
    \bm{m}^{(l)}_n = f_{\theta}(\{\bm{h}^{(l-1)}_{n} | n \in \mathcal{N}(a)\}),  
\end{align}
where $\mathcal{N}(a)$ is the set of neighbors of node $a$. Then, node $a$ updates its embedding by aggregating the received messages:
\begin{align}
    \bm{h}^{(l)}_a = \mathcal{A}(\bm{m}^{(l)}_{n} | n \in \mathcal{N}(a)\}).
\end{align}
After $L$ iterations of propagation and aggregation, each node's representation incorporates both topological and attribute information from its $L$-hop neighbours in the graph, thereby capturing spatial dependencies among nodes.

\subsection{Mamba}
The Mamba model is a recently proposed neural architecture designed to efficiently model long-range temporal dependencies while maintaining linear computational complexity, offering an alternative to attention-based mechanisms, such as Transformer. It is built upon the SSM framework, which models sequences as continuous-time dynamical systems. Mamba leverages the state-space representation and introduces a selective mechanism that dynamically controls information retention and forgetting at each time step. In the continuous-time SSM, the evolution of the hidden state $\bm{z}(t)$ and the output $\bm{O}(t)$ are defined as:
\begin{align}
  \label{ssm}
    \bm{z}^{'}(t) =  B\bm{z}(t) + C\bm{H}(t),\,
    \bm{O}(t) = D\bm{z}(t),
\end{align}
where $\bm{H}(t)$ is the input sequence, and $B$, $C$, $D$ are learnable parameter matrices. For discrete input sequences, the continuous-time SSM in Eq.~(\ref{ssm}) is discretized using the zero-order hold method~\cite{Ding2025DyGMamba}, yielding the following discrete formulation:
\begin{align}
   \label{ssm2}
    \bm{z}(t) = \overline{B}\bm{z}(t-1) + \overline{C}\bm{H}(t), \, \bm{O}(t) = D \bm{z}(t),
\end{align}
 where $\overline{B} = \exp{(\Delta B)}$, $\overline{C} = (\Delta B)^{-1}(\exp{(\Delta B)}-I)(\Delta C)$, $\Delta$ is a specified sampling timescale for the discretization. Mamba combines the dynamical representation power of state-space models with adaptive selection mechanisms to learn complex temporal dependencies.

\section{Spatiotemporal Trust Evaluation via Customized GNN-Mamba}


Leveraging the strengths of GNNs and Mamba, we propose a customized spatiotemporal trust evaluation model, GM, for collaborator selection. It comprises five key steps: construction of historical collaboration-based graph snapshots, GNN-aided spatial trust fusion, Mamba-aided temporal trust fusion, historical trust calculation, and task-specific resource trust evaluation.

\subsection{Construction of Historical Collaboration-based Graph Snapshots}

To infer device trust from historical collaborations, we adopt a trust-centric modeling approach that treats collaboration outcomes as evidence contributing to trust accumulation over time. Rather than directly aggregating all past collaborations, trust evidence is progressively collected within successive observation windows to reflect its temporal evolution. Let the observation period be divided into $S$ consecutive intervals $\{t_1,\dots,t_S\}$. Within each interval $t_s$, only the collaboration records observed within this interval are considered when inferring trust. Then, a directed trust graph $G(t_s)$ is derived, where nodes correspond to devices involved in task execution during $t_s$, and directed edges represent trust attribution relationships induced by task assistance. Specifically, when device $a_j$ participates in executing tasks delegated by device $a_i$, an edge from $a_i$ to $a_j$ is introduced, indicating that $a_i$ assigns trust to $a_j$ based on the observed collaboration outcomes. The strength of this trust attribution is quantified by aggregating $N_{(a_i,a_j)}$ collaboration outcomes between device $a_i$ and device $a_j$ during $t_s$ as: 
\begin{align}
    T_{(a_i, a_j)}^{\text{col}} = \frac{1}{N_{(a_i,a_j)}} \sum_{n=1}^{N_{(a_i,a_j)}} 
    \left(\alpha_1 (1 - p_n^{\text{loss}}) + \alpha_2 p_n^{\text{out}}\right),
\end{align}
where $p_n^{\text{loss}}$ characterizes the communication reliability (packet loss rate) of the $n$-th collaboration, and $p_n^{\text{out}}$ reflects its task execution outcome. The weighting coefficients $\alpha_1$ and $\alpha_2$ control the relative contributions of communication and computation performance, with $\alpha_1+\alpha_2=1$. Each snapshot graph $G(t_s)$ thus captures trust evidence accumulated within interval $t_s$. Ordering the snapshots according to their temporal indices yields a time-evolving graph sequence $\{G(t_1),\dots,G(t_S)\}$, which serves as the basis for subsequent spatiotemporal trust modeling.

\subsection{GNN-Aided Trust Fusion in Spatial Dimension}
In collaborative systems, trust is inherently role-dependent: a device evaluates others as a \emph{trustor} while simultaneously being evaluated as a \emph{trustee}. To explicitly capture this asymmetry, we model trust as a role-centric representation learning problem over historical collaboration graphs, where each device maintains distinct embeddings corresponding to its trustor and trustee roles. Given a historical collaboration graph $G(t_s)$ at time slot $t_s$, directed edges represent task-oriented collaborations between devices, naturally inducing asymmetric trust relationships. Rather than directly propagating trust values across the graph, we first construct role-specific latent representations that reflect how trust evidence is accumulated and interpreted under different roles.

\subsubsection{Trust fusion under the trustee role}

When device $a_j$ acts as a trustee, its trustworthiness is inferred from the collective feedback of devices that have previously relied on it for task execution. These devices correspond to device $a_j$'s one-hop in-degree neighbors $\mathcal{N}_{\mathrm{in}}(a_j)$ in $G(t_s)$. For neighbor $a_i \in \mathcal{N}_{\mathrm{in}}(a_j)$, its associated trust evidence is mapped into a latent trust signal:
\begin{align}
    \bm{u}_{a_j \gets a_i}^{\text{tr}} = \mathbf{P}^{\text{tr}}_{a_{j}\gets a_{i}} \bm{\psi}_{a_j \gets a_i},
\end{align}
where $\bm{\psi}_{a_j \gets a_i}$ represents the binary-encoded trust value $T^{\text{col}}_{(a_i,a_j)}$, and $\mathbf{P}^{\text{tr}}_{a_j\gets a_i} \in \mathbb{R}^{d_a \times d_T}$ is a learnable projection matrix. A trainable temporal embedding $\bm{\tau}_{t_s} \in \mathbb{R}^{d_a}$ encodes the current time slot $t_s$~\cite{YuSunDuLv23_DyGFormer}. 
The influence of device $a_i$ on the trustworthiness of device $a_j$ is then captured by:
\begin{align}
    \bm{m}_{a_j \gets a_i}^{\text{tr}} = f_\theta(\bm{h}_{a_i}, \bm{u}_{a_j \gets a_i}^{\text{tr}}, \bm{\tau}_{t_s}),
\end{align}
where $\bm{h}_{a_i} \in \mathbb{R}^{d_a}$ represents the latent state of device $a_i$, which is generated via node2vec, and $f_\theta(\cdot)$ is a learnable message function. The trustee representation of $a_j$ is obtained by aggregating these role-conditioned signals:
\begin{align}
    \bm{h}_{a_j}^{\text{tr}} = \mathcal{A}\Big(\{\bm{m}_{a_j \gets a_i}^{\text{tr}} \mid a_i \in \mathcal{N}_{\text{in}}(a_j)\}\Big).
\end{align}

\subsubsection{Trust fusion under the trustor role}
When device $a_j$ acts as a trustor, it expresses trust toward selected collaborators. Let $\mathcal{N}_{\mathrm{out}}(a_j)$ represent the set of one-hop out-degree neighbors of device $a_j$ in $G(t_s)$. Device $a_j$ aggregates trust information propagated by its one-hop out-degree neighbours. This process is formulated as: \begin{align}
    \bm{u}_{a_j\to a_r}^{\text{to}} &= \mathbf{P}^{\text{to}}_{a_j\to a_r} \bm{\psi}_{a_j \to a_r}, a_r \in \mathcal{N}_{\mathrm{out}}(a_j)\\
    \bm{m}_{a_j\to a_r}^{\text{to}} &= f_\theta(\bm{h}_{a_r}, \bm{u}_{a_j\to a_r}^{\text{to}}, \bm{\tau}_{t_s}),
\end{align}
where $\mathbf{P}^{\text{to}}_{a_j\to a_r}$ is a learnable projection matrix, $\bm{\psi}_{a_j \to a_r}$ denotes the encoded trust value $T^{\text{col}}_{(a_j,a_r)}$. Aggregating these signals from neighbours yields a trustor-specific representation:
\vspace{-0.1 in}
\begin{align}
    \bm{h}_{a_j}^{\text{to}} = \mathcal{A}\Big(\{\bm{m}_{a_j\to a_r}^{\text{to}} \mid a_r \in \mathcal{N}_{\text{out}}(a_j)\}\Big).
\end{align}

\subsubsection{Role fusion}
The final trust state of device $a_j$ at time $t_s$ is obtained by fusing its role-specific representations:
\begin{align}
    \bm{h}_{a_j}(t_s) = \sigma \Big( \mathbf{W}_{\mathrm{fuse}} 
    ( \bm{h}_{a_j}^{\mathrm{tr}} \parallel \bm{h}_{a_j}^{\mathrm{to}} ) 
    + \mathbf{b}_{\mathrm{fuse}} \Big),
\end{align}
where $\sigma(\cdot)$ denotes a nonlinear activation. Stacking multiple such layers enables progressive refinement of trust representations by incorporating multi-hop dependencies.

Over a sequence of time slots $\mathbb{T}=\{t_1,\dots,t_S\}$, the historical trust trajectory of device $a_j$ is represented as:
\begin{align}
    \bm{H}_{a_j} = [\bm{h}_{a_j}(t_1); \bm{h}_{a_j}(t_2); \dots; \bm{h}_{a_j}(t_S)].
\end{align}
This role-centric, temporally-aware embedding encodes how device $a_j$’s trustworthiness evolves over time. The same procedure is applied to all devices to generate role-aware embeddings that reflect spatial trust dependencies.

\subsection{Mamba-Aided Trust Fusion in Temporal Dimension}

To effectively model the temporal evolution of trust for each device $a_j$, we perform temporal fusion over its embedding $\bm{H}_{a_j}$. Since $\bm{H}_{a_j}$ captures trust information across multiple time slots, it is important to account for both short-term fluctuations and long-term trends. Additionally, considering dependencies from both previous and future time steps allows the model to extract more informative temporal features.

To achieve this, we employ a Mamba-assisted temporal fusion approach, which consists of $M$ stacked Mamba blocks. Each Mamba block combines a convolutional operation for capturing local temporal correlations with a discrete SSM module that models long-range dependencies. In each Mamba block,  device $a_j$'s embedding $\bm{H}_{a_j}$ first passes through a convolutional layer, which is given by:
\begin{align}
   \Check{\bm{H}}_{a_j} &= \text{Linear}(\bm{H}_{a_j}), \\
   \widetilde{\bm{H}}_{a_j}  &= \text{SiLU}(\text{Conv1D}(\Check{\bm{H}}_{a_j})),
\end{align}
where $\text{Linear}(\cdot)$ is the linear layer, $\text{Conv1D}(\cdot)$ is the 1D convolution layer, and $\text{SiLU}(\cdot)$ is the activation function. Next, the processed embedding $\widetilde{\bm{H}}_{a_j}$ is input into the discrete SSM module:
\vspace{-0.1 in}
\begin{align}
   \bm{O}_{\text{SSM}} = \text{SSM}_{(\overline{B}, \overline{C}, D)}(\widetilde{\bm{H}}_{a_j}),
\end{align}
which recursively integrates information across the sequence, ensuring that each state captures both current inputs and accumulated historical context. The detailed steps of the Mamba block are provided in Algorithm~\ref{mambablock}. To obtain a fixed-size representation suitable for downstream evaluation, we apply a max-pooling operation over the generated $\bm{O}_{a_j}$, producing the final embedding:
\begin{align}
\widehat{\bm{O}}_{a_j} = \text{maxpooling}(\bm{O}_{a_j}) \in \mathbb{R}^{d_a}.
\end{align}

\begin{figure}[t!]
\centering
\includegraphics[scale=1]{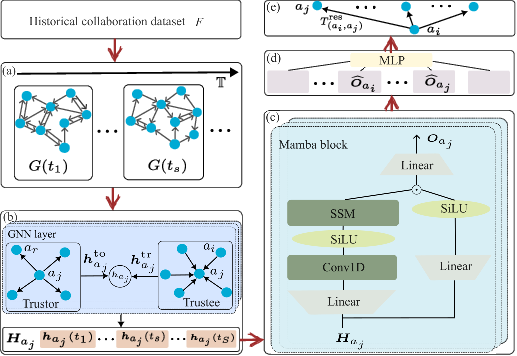}
\label{gnn}
\end{figure}
\begin{algorithm}[t]
\caption{Mamba block}
\begin{algorithmic}[1]  
\State \textbf{Input}: $\bm{H}_{a_j}$
\State \textbf{Output}: $\bm{O}_{a_j}$   
\State  $\Check{\bm{H}}_{a_j} \gets \text{Linear}(\bm{H}_{a_j})$
\State $\widetilde{\bm{H}}_{a_j}  \gets \text{SiLU}(\text{Conv1D}(\Check{\bm{H}}_{a_j}))$ 
\State initialize $B$ as a dialogue matrix 
\State $C \gets \text{Linear}(\widetilde{\bm{H}}_{a_j})$
\State  $D \gets \text{Linear}(\widetilde{\bm{H}}_{a_j})$
\State $\Delta \gets \text{Softplus}(\text{Linear}(\widetilde{\bm{H}}_{a_j}))$
\State $\overline{B} \gets \text{Discrete}(\Delta, B)$
\State $\overline{C} \gets \text{Discrete}(\Delta, B, C,)$
\State $\bm{O}_{\text{SSM}} \gets \text{SSM}_{(\overline{B}, \overline{C}, D)}(\widetilde{\bm{H}}_{a_j})$ 
\State $\bm{O}_{a_j} \gets  \text{Linear}(\bm{O}_{\text{SSM}} \odot \text{SiLU}(\text{Linear}(\bm{H}_{a_j})))$
\end{algorithmic}
\label{mambablock}
\end{algorithm}

\subsection{Historical Trust Calculation}
By fusing information across both spatial and temporal dimensions, we obtain embeddings for all devices that are rich in trust-related information. Given any device pair, e.g., $a_i$ and $a_j$, we concatenate the pooled embeddings $\widehat{\bm{O}}_{a_i}$ and $\widehat{\bm{O}}_{a_j}$ and feed them into a Multi-Layer Perceptron (MLP). The MLP outputs a trust score vector $\bm{T}^{\text{his}}_{(a_i,a_j)}$, from which the maximum value is selected to present the final trust assessment: \begin{align}
    T^{\text{his}}_{(a_i,a_j)} &= \max(\text{MLP}(\widehat{\bm{O}}_{a_i} \parallel \widehat{\bm{O}}_{a_j})),
\end{align}
To train the GNN and Mamba models, we define the objective function as the cross-entropy loss between the computed trust values and the ground-truth trust values:
\begin{align}
    \mathcal{L} = \text{cross\_entropy}(\bm{T}^{\text{col}}, \bm{T}^{\text{his}}),
\end{align}
where $\bm{T}^{\text{his}}$ is the set of computed values, and $\bm{T}^{\text{col}}$ is the set of ground-truth values.

\subsection{Task-Specific Resource Trust Evaluation}

After obtaining the historical trustworthiness of all potential collaborators, the task owner $a_i$ further evaluates the resource trustworthiness of each collaborator for the task $\beta$. The evaluation considers whether the task can be completed within the task owner’s maximum tolerable time. It is worth noting that the criteria for resource trustworthiness are extensible, allowing incorporation of additional metrics if necessary. The estimated task completion time for a potential collaborator includes both the transmission time from the task owner and the computation time required by the collaborator. By collecting communication-related resources, the task transmission time from the task owner $a_i$ to collaborator $a_j$ is estimated as:
\begin{align}
    t^{\text{tra}}_{(a_i,a_j)} ={\beta^{\text{size}}}/{r_{(a_i,a_j)}},
\end{align}
where $\beta^{\text{size}}$ is the task size, and $r_{(a_i,a_j)}$ represents the achievable transmission rate from $a_i$ to $a_j$. $r_{(a_i,a_j)}$ can be calculated as $W^{\text{band}}\log_2(1 + (a_i^{\text{pow}} g_{(a_i,a_j)})/N_0)$, where $W^{\text{band}}$ is the channel bandwidth, $a_i^{\text{pow}}$ is the transmission power of $a_i$, $g_{(a_i,a_j)}$ denotes the channel gain between devices $a_i$ and $a_j$, and $N_0$ is the noise power. By collecting computation-related resources, the task computation time required by collaborator $a_j$ is estimated as~\cite{11096939}:
\begin{align}
    t^{\text{com}}_{a_j} = (\beta^{\text{size}}\beta^{\text{des}})/a_j^{\text{CPU}},
\end{align}
where $a_j^{\text{CPU}}$ is the available CPU frequency of collaborator $a_j$. If the sum of task transmission time and task computation time does not exceed the task owner's maximum tolerable time, i.e., $t^{\text{tra}}_{(a_i,a_j)} + t^{\text{com}}_{a_j} \leq \beta^{\text{time}}$, the resources of collaborator $a_j$ are considered trusted, $T^{\text{res}}_{(a_i,a_j)} = 1$; otherwise, its resources are deemed untrusted, $T^{\text{res}}_{(a_i,a_j)} = 0$. Based on each collaborator $a_j$'s historical trust $T^{\text{his}}_{(a_i,a_j)}$ and task-specific resource trust $T^{\text{res}}_{(a_i,a_j)}$, the task owner $a_i$ selects the collaborator with the highest trust value $T_{(a_i,a_j)}$ as the final collaborator.

\section{Experimental Analysis}

\subsection{Experimental Setup}
\subsubsection{System configuration}
We implement the proposed GM model in a wireless system using the discrete-event network simulator NS-3 with Python bindings~\cite{ns3website}. NS-3 allows accurate simulation of realistic network behavior and supports distributed computation modeling. Key device and network parameters are summarized in Table~\ref{tab:device_params}.

\begin{table}[t!]
\centering
\caption{Device and network parameters}
\label{tab:device_params}
\begin{tabular}{l c}
\hline
\text{Parameter} & \text{Value} \\
\hline
\hline
Number of devices & 500 \\
Transmission power & 100 mW \\
CPU frequency & Randomly selected from $\{2,4,6\}$ GHz \\
Channel bandwidth & 5 MHz \\
Noise power & -80 dBm \\
Trust weights & $\alpha_1 = 0.6$, $\alpha_2 = 0.4$ \\
\hline
\end{tabular}
\end{table}

\subsubsection{Task modeling}
Two types of tasks are modeled: face recognition and virus scanning. Task characteristics used in the experiments are summarized in Table~\ref{tab:task_params}. A total of 8,000 tasks are executed sequentially, and each device's performance is recorded. Ground-truth trust labels are generated from historical device performance following~\cite{Favour2025Benchmarking}.

\begin{table}[t!]
\centering
\caption{Task parameters}
\label{tab:task_params}
\begin{tabular}{l c c}
\hline
 & \text{Face recognition} & \text{Virus scanning} \\
\hline
Task size (MB) & 5 & 5 \\
Processing density (cycles/bit) & 2,339~\cite{11296817} & 32,946~\cite{11296817} \\
Max tolerable time (s) & 80 & 700 \\
\hline
\end{tabular}
\end{table}

\subsubsection{Model and training strategy}
The GM model's embedding, GNN, Mamba, and training hyperparameters are summarized in Table~\ref{tab:training_params}. The dataset is split into 80\% training and 20\% testing subsets. Five-fold cross-validation and early stopping are applied to prevent overfitting. All experiments are performed on the Lambda Vector workstation.  Without specification, the reported experiment results correspond to a learning rate of $10^{-2}$, an $L_2$ regularization coefficient of $10^{-5}$, and a dropout rate of $0$. 

\begin{table}[t!]
\centering
\caption{Model and training parameters}
\label{tab:training_params}
\begin{tabular}{l c}
\hline
\text{Parameter} & \text{Value} \\
\hline
Embedding dimension & 128 \\
GNN layers & 3 (outputs: 32, 64, 32) \\
Mamba layers & 4  \\
Learning rate & $\{10^{-1}, 10^{-2}, 10^{-3}, 10^{-4}\}$ \\
$L_2$ regularization & $\{10^{-5},10^{-4}\}$ \\
Dropout rate & $\{0,0.1,0.3,0.5,0.8\}$ \\
\hline
\end{tabular}
\end{table}

\begin{table}[!t] 
    \footnotesize  
    \centering
    \renewcommand{\arraystretch}{1.3}
    \caption{Comparative Analysis of RMSE and MAE}
    \label{rmse}
    \begin{tabular}{m{2.5cm}<{\centering}|m{2cm}<{\centering}|m{2cm}<{\centering}}
        \hline 
         & RMSE & MAE \\
        \hline
        \hline
        QS-Trust~\cite{Najib2024QSTrust} & 0.171 $\pm$ 0.005 & 0.132 $\pm$ 0.004  \\
        \hline
         Medley (GNN)~\cite{9488814} & 0.131 $\pm$ 0.004  &   0.109 $\pm$ 0.003 \\
        \hline
        LSTM~\cite{Alghofaili2022TrustManagement} &  0.122 $\pm$ 0.005 & 0.102 $\pm$ 0.002   \\
        \hline
        \textbf{GM} &  0.104 $\pm$ 0.004 & 0.083 $\pm$ 0.002\\
        \hline
    \end{tabular}
\end{table}

\subsection{Comparison of Historical Trust Evaluation}

We first evaluate the performance of the proposed GM model in historical trust assessment. Two widely used metrics–Root Mean Square Error (RMSE) and Mean Absolute Error (MAE)–are employed to evaluate the accuracy of trust assessment. Lower values of RMSE and MAE indicate higher accuracy. The comparison results with baseline methods are presented in Table~\ref{rmse}. The proposed GM method achieves the lowest RMSE and MAE values, demonstrating high accuracy. This improvement arises from GM’s ability to capture and integrate trust information across both spatial and temporal dimensions. In contrast, LSTM captures only temporal dependencies, Medley considers only spatial dependencies, and the rule-based QS-Trust neglects both, resulting in the poorest performance.

\subsection{Comparison of Task-Specific Resource Trust Evaluation}
\begin{figure}[!t]
      \centering
      \subfigure[]{\includegraphics[scale=0.66]{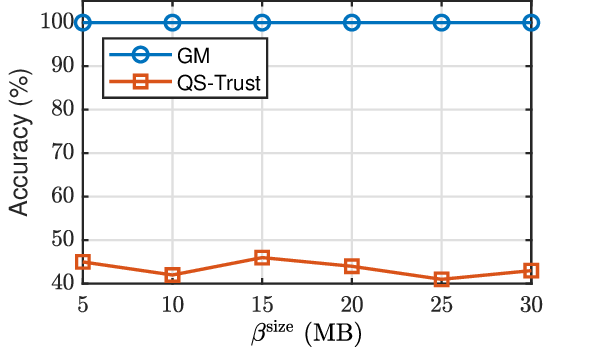}} \\
      \hspace{-0.01 in}\subfigure[]{\includegraphics[scale=0.67]{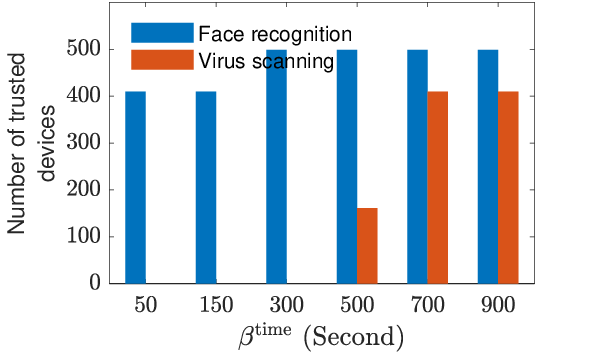}}
      \caption{Comparative analysis of task-specific resource trust. (a) The proposed GM achieves 100\% accuracy in task-specific resource trust evaluation across different task sizes. (b) As the task owner's maximum tolerable time increases, more devices are deemed resource-trustworthy. The face recognition task includes more trusted devices than the virus scanning task, as it requires a shorter processing time.}
     \label{task}
\end{figure}
\begin{figure}[!t]
\centering
\includegraphics[scale=0.64]{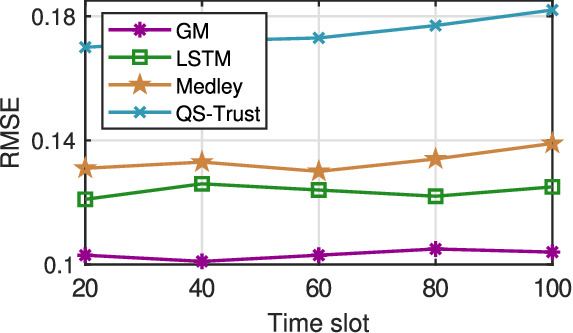}
\caption{The RMSE comparison of $T_{(a_i,a_j)}$ across different time slots demonstrates that the proposed GM model exhibits the smallest fluctuations.}
\label{timeslot}
\end{figure}

We further investigate the impact of task attributes on task-specific resource trustworthiness $T^{\text{res}}_{(a_i,a_j)}$. As shown in Fig.~\ref{task}~(a), the GM model consistently identifies trustworthy devices with 100\% accuracy across different sizes of the face recognition task. In contrast, QS-Trust achieves less than 50\% accuracy, owing to its generic resource evaluation approach that ignores task-specific features. We set the size of both tasks to $10 \text{ MB}$ and vary the maximum tolerable time ($\beta^{\text{time}}$) to observe the resulting number of resource-trustworthy devices. As shown in Fig. $\text{\ref{task}}$ (b), the number of resource-trusted devices gradually increases for both tasks as $\beta^{\text{time}}$ increases. This trend is expected, as a longer tolerable time allows a larger pool of devices to meet the deadline constraint. In addition, the number of resource-trusted devices for the virus scanning task is significantly lower than that for the face recognition task. This is because the virus scanning task has a higher processing density, which consequently leads to longer task execution times, thereby disqualifying more devices from meeting the maximum tolerable time. The experimental results demonstrate that the proposed GM model can effectively identify devices that are trustworthy for specific tasks.

\subsection{Trust Evaluation Comparison Across Time Slots}

Fig.~\ref{timeslot} compares the RMSE of the evaluated trust values $T_{(a_i,a_j)}$ produced by different methods as the number of time slots increases. The proposed GM consistently achieves the lowest RMSE across all time slots. Moreover, it exhibits the smallest fluctuations, demonstrating strong adaptability and robustness to dynamic data with spatiotemporal dependencies.

\section{Conclusion}
This study has investigated the accurate evaluation of device trustworthiness by integrating multiple trust-related perspectives to enable efficient collaborator selection. The customized GM model is proposed, where the GNN model performs spatial fusion of trust information to capture inter-device trust dependencies, while the Mamba model realizes temporal fusion to characterize the evolution of trust over time. In addition, the GM model integrates task-specific resource trust evaluation to comprehensively reflect the trustworthiness of devices across different task scenarios. This work provides a new perspective for device trust modeling in complex spatiotemporal environments and lays a foundation for achieving more reliable and intelligent collaboration.

\vspace{0.07 in}
\footnotesize

\end{document}